\documentclass[prd,amssymb,twocolumn,10pt]{revtex4}
\usepackage{amsmath,amsfonts,amssymb}
\usepackage{verbatim,graphicx,float}  

\begin{document}

\title{Tensor power spectrum with holonomy corrections in LQC}

\author{Jakub Mielczarek}
\email{jakub.mielczarek@uj.edu.pl}
\affiliation{Astronomical Observatory, Jagiellonian University, 30-244
Cracow, Orla 171, Poland}

\begin{abstract}
In this paper we consider tensor perturbations produced at a bounce phase 
in presence of the holonomy corrections. Here bounce phase and holonomy 
corrections originate from Loop Quantum Cosmology. We re-derive formulas 
for the corrections for the model with a scalar field content. Background 
dynamics with a free scalar field and multi-fluid potential are considered. 
Since the considerations are semi-classical effects of quantum fluctuations
of the background dynamics are not taken into account. Quantum and classical 
back-reaction effects are also neglected. To find spectrum of the gravitational 
waves both analytical approximations as well as numerical investigations 
were performed. We have found analytical solutions on super-horizontal and 
sub-horizontal regimes and derived corresponding tensor power spectra. Also 
occupation number $n_{\bf k}$ and parameter $\Omega_{\text{gw}}$ were derived 
in sub-horizontal limit, leading to its extremely low present value. Final 
results are numerical power spectra of the gravitational waves produced in
presence of quantum holonomy corrections. The obtained spectrum has two 
UV and IR branches where $\mathcal{P}_T \propto k^2$, however with the different 
prefactors. Spectrum connecting these regions is in the form of oscillations.
We have found good agreement between numerical spectrum and this obtained 
from the analytical model. Obtained spectrum can be directly applied
as initial conditions for the inflationary modes. We mention possible resulting
observational features of the CMB in particular B-type polarization.
\end{abstract}

\maketitle

\section{Introduction} \label{sec:intro}

In the Minkowski background free gravitational waves fulfil the wave equation 
$(\partial_t^2-\nabla^2 )h_{\mu\nu}=0$. Solutions of this equation are plane waves 
$h_{\mu\nu}=P_{\mu\nu}e^{i({\bf k \cdot x}-|{\bf k}| t)}$, here $P_{\mu\nu}$ is polarization tensor.
However when cosmological expansion  is turned on (we assume flat FRW background here) additional 
term appears and the equation of motion is modified to $(\partial_t^2+3H\partial_t-\nabla^2 )h_{\mu\nu}=0$,
where $H$ is Hubble parameter. We see that cosmological term acts as effective friction.
When Universe undergoes expansion then $H>0$ and gravitational waves are damped.
This situation corresponds to the present stage of the evolution.
However when Universe is in contracting phase $H<0$ friction term becomes negative 
and gravitational waves are amplified. Such a phase of contraction is 
general prediction of Loop Quantum Cosmology (LQC)\cite{Bojowald:2008zzb}. 
Contracting phase appears also in the string theory 
based theories of the Universe \cite{Shtanov:2002mb,Gasperini:2002bn} and many others. 
However in the present paper we concentrate on the LQC 
models where contracting and expanding regimes are joined by the phase of bounce \cite{Ashtekar:2006rx}.
In the last years perturbations during the bounce phase were studied extensively.
Recent review on this issue can be found in \cite{Wands:2008tv}. In particular 
perturbations in the Quintom Bounce were studied in \cite{Cai:2008ed,Cai:2007zv}.

In LQC many physical results can be traced in semi-classical approximation.
In particular dynamics of the Universe can be recovered from the quantum 
corrected Friedmann equation \cite{Singh:2006im,Mielczarek:2008zv}. Similar 
approach can be also applied to describe quantum gravity effects on perturbations
\cite{Bojowald:2006tm}, in particular to gravitational waves. It is however worth
to stress that such an approach is rather heuristic and results obtained have to 
be verified by the purely quantum considerations. In particular it has not 
been proved yet whether phase of bounce is generally realized for the inhomogeneous
loop cosmologies. However some recent studies show that in case of loop quantized 
inhomogeneous Gowdy spacetime, singularity is avoided \cite{Brizuela:2009nk}. In our 
approach inhomogeneities are treated perturbatively and we neglect their back-reaction 
on the background dynamics. However in the more detailed studies these effects have to be 
also taken into account. In the semi-classical approach applied here, quantum gravity 
effects are introduced by the corrections to the classical equations of motion. 
For tensor modes in LQC these effects were preliminary studied 
in Ref. \cite{Mielczarek:2007zy,Mielczarek:2007wc}. Later improved
approach was developed \cite{Bojowald:2007cd} introducing holonomy corrections.
Results of this paper are a backbone of our investigations. In this paper we assume 
that these corrections are valid during the whole evolution.  
Some preliminary studies of influence of the holonomy corrections 
for the gravitational waves production have been done \cite{Barrau:2008hi,Mielczarek:2008pf,Grain:2009kw}.
However in that papers effects of the corrections to the source term were neglected.
While in classical approach this term vanishes (within linear regime) in quantum 
regime it does contribute. In the present paper we improve these studies including a
source term. 

Besides the holonomy corrections also inverse volume corrections are predicted in the framework of LQC.
Effects of inverse volume corrections on gravitational waves were recently studied in
Ref. \cite{Copeland:2008kz,Calcagni:2008ig}. However in the flat FRW background
inverse volume corrections exhibit fiducial cell dependence. This makes those 
effects harder to interpret. However, in the curved backgrounds this problem disappears.
Since holonomy and inverse volume effects differ qualitatively, they should be 
studied separately. In this paper we follow this line of reasoning. We consider consistent model 
where holonomy corrections influence both background and perturbations parts.

The organisation of the text is the following. In section \ref{sec:HoloCorr}  
we introduce the equation for tensor modes with holonomy corrections.
Then in section \ref{sec:Backgroun} we define background dynamics.
We consider both the model with free scalar field and with multi-fluid potential. 
Subsequently  in section \ref{sec:Analytic} and section \ref{sec:Numerical}
we investigate analytically and numerically the evolution of the tensor modes.
Effects of holonomy corrections are investigated. With use of numerical computations 
we calculate power spectra of the gravitational waves.
In section \ref{sec:Summary} we summarize the results. 
Finally in Appendix we introduce gravitational waves in LQC framework,
derive particular form of the holonomy corrections and explain the employed notation.

\section{Gravitational waves with holonomy corrections} \label{sec:HoloCorr}

Equation for tensor modes with LQC holonomy corrections derived in \cite{Bojowald:2007cd}  is given by
\begin{equation}
\frac{d^2}{d\eta^2}h^i_a + 2\bar{k}\frac{d}{d\eta}h^i_a-\nabla^2h^i_a+
T_Qh^i_a = 16 \pi G \Pi^i_{Qa}, \label{TensorEq1}
\end{equation}
where 
\begin{eqnarray}
T_Q &=&  -2\left(\frac{\bar{p}}{\bar{\mu}} \frac{\partial \bar{\mu}}{\partial \bar{p}}  \right)
\bar{\mu}^2\gamma^2\left( \frac{\sin \bar{\mu}\gamma \bar{k} }{ \bar{\mu}\gamma} \right)^4,  \label{TQ}      \\
\Pi^i_{Qa} &=& \left[ \frac{1}{3V_0} \frac{\partial \bar{H}_{\text{m}}}{\partial \bar{p}} 
\left( \frac{\delta E^c_j\delta^j_a \delta^i_c}{\bar{p}} \right)\cos 2\bar{\mu}\gamma \bar{k}
+\frac{\delta H_m}{\delta(\delta E^a_i)}  \right].  \label{PiQ}
\end{eqnarray}
For details and explanation of the employed notation we send to Appendix. To derive specific form of 
the functions $T_Q $ and $\Pi^i_{Qa}$, matter content must be defined.
In this paper we consider models with a scalar field. We consider both free and 
self-interacting fields. In that case matter Hamiltonian is up to the second order 
\begin{equation}
{H}_{m}=\bar{H}_{m} + \frac{1}{4}\int_{\Sigma} d^3{\bf x} \, \frac{\bar{N}}{\sqrt{\bar{p}}} 
 \left( \frac{1}{2}\frac{\pi^2_{{\phi}}}{\bar{p}^{3} }-V({\phi}) \right)\delta^i_a \delta E^a_j \delta^j_b \delta E^b_i. 
\end{equation}
where homogeneous part is given by 
\begin{equation}
\bar{H}_{m} = V_0\bar{N} \bar{p}^{3/2}\left( \frac{1}{2}\frac{\pi^2_{{\phi}}}{\bar{p}^{3} }+V({\phi}) \right).
\end{equation}
Here integration was constrained to fiducial volume $V_0$. 
Further physical results do not depend of this quantity.  

Energy density can be now defined as  
\begin{equation}
\rho := \frac{1}{V_0\bar{p}^{3/2}} \frac{\partial \bar{H}_{m}}{\partial \bar{N}}.
\end{equation}

When matter content is defined one can derive particular form 
of the functions (\ref{TQ}) and  (\ref{PiQ}).  
Expressions for the quantum holonomy corrections simplify to 
\begin{eqnarray}
T_Q        &=& \frac{8\pi G}{3} \frac{\bar{p}\rho^2}{\rho_c},  \\
\Pi^i_{Qa} &=&  \Pi_Qh^i_a = \frac{1}{2}\bar{p} \frac{\rho}{\rho_c} \left(2V-\rho\right)h^i_a.  \label{PiQSF}       
\end{eqnarray}
These expressions were first derived in Ref. \cite{Copeland:2008kz}.
However we have found a discrepancy between the expression for $\Pi^i_{Qa}$ derived 
here and this found in  Ref. \cite{Copeland:2008kz}. To approve the result 
presented here we show intermediate steps of derivation in Appendix.  
The difference is $1/3$ factor inside the bracket.        
To derive these corrections we have applied so called $\bar{\mu}$ scheme of 
quantisation. Namely we used $\bar{\mu}=\sqrt{\Delta/\bar{p}}$ where 
$\Delta = 2\sqrt{3}\pi \gamma l^2_{\text{Pl}}$. It is well motivated to 
use this particular form of the function \cite{Corichi:2008zb}. However, other choices
are in principle also permitted. In this paper we consider only 
$\bar{\mu}$ scheme, which seems to be the best motivated.

Now equation for the tensor modes (\ref{TensorEq1}) simplifies to 
\begin{equation}
\frac{d^2}{d\eta^2}h^i_a + 2\bar{k}\frac{d}{d\eta}h^i_a-\nabla^2h^i_a+
\tilde{T}_Qh^i_a = 0 \label{GWEq1}
\end{equation}
where we have defined the total holonomy correction 
\begin{equation}
\tilde{T}_Q = T_Q-16 \pi G \Pi_Q  = 16 \pi G \bar{p} \frac{\rho}{\rho_c}\left(
\frac{2}{3}\rho-V \right).       
\end{equation}
Therefore also source term correction has been included. This is in contrast 
with the analysis performed in \cite{Barrau:2008hi,Mielczarek:2008pf,Grain:2009kw},
where this influence was neglected. In the classical theory in fact this term vanish 
in the linear order. Therefore when fluctuations of vacuum are considered, 
higher order term can be set to zero.  However since, due to quantum corrections,
source term contribute linearly, there is no reason to neglect this term. 
Therefore in the present paper we take it into account.

We introduce new common variable
\begin{equation}
u=\frac{a h_{\oplus}}{\sqrt{16\pi G}}=\frac{a h_{\otimes}}{\sqrt{16\pi G}},  
\end{equation}
where $h^1_1=-h^2_2=h_{\oplus}$, $h^1_2=h^2_1=h_{\otimes}$ and $a=\sqrt{\bar{p}}$.
Then performing the Fourier transform
\begin{equation}
u(\eta,{\bf x}) = \int \frac{d^3{\bf k}}{(2\pi)^3}  u(\eta,{\bf k}) e^{i{\bf k}\cdot {\bf x}}, 
\end{equation}
one can rewrite the equation (\ref{GWEq1}) in the form
\begin{equation}
\frac{d^2}{d\eta^2} u(\eta,{\bf k}) +[k^2 +m^2_{\text{eff}}] u(\eta,{\bf k}) = 0, \label{modeeq}
\end{equation}
where $k^2={\bf k}\cdot {\bf k}$ and 
\begin{equation}
m^2_{\text{eff}}  = \tilde{T}_{Q} - \frac{a^{''}}{a}.  
\end{equation}
In this paper we aim to solve equation (\ref{modeeq}). 
However first we must specify the background dynamics.

\section{Background dynamics} \label{sec:Backgroun}

Background dynamics is governed by the effective Friedmann equation
\begin{equation}
\left( \frac{1}{2\bar{p}}  \frac{d\bar{p}}{dt}\right)^2=\frac{\kappa}{3}\rho \left(1-\frac{\rho}{\rho_c}\right),
\end{equation}
where 
\begin{equation}
\rho_c = \frac{\sqrt{3}}{16\pi^2 \gamma^3 l^4_{\text{Pl}}}
\end{equation}
is critical energy density. This equation can be derived combining the
Hamilton equation $\dot{\bar{p}}=\{\bar{p}, \bar{H}_{m}+\bar{H}_{G}^{\text{phen}}\}$ with 
the scalar constraint $\bar{H}_{m}+\bar{H}_{G}^{\text{phen}} = 0$.

Evolution of the scalar field component is governed
by the Hamilton equations
\begin{eqnarray}
\dot{\phi} &=& \{\phi,\bar{H}_m  \}=  \bar{p}^{-3/2}  \pi_{\phi}, \\
\dot{\pi_\phi} &=& \{\pi_\phi, \bar{H}_{m}  \}=-\bar{p}^{3/2}\frac{dV}{d\phi}. 
\end{eqnarray}

Energy density and pressure of the homogeneous scalar field are expressed as follows 
\begin{eqnarray}
\rho_{\phi} &=& \frac{1}{2}\dot{\phi}^2+V(\phi),  \\
p_{\phi} &=& \frac{1}{2}\dot{\phi}^2 - V(\phi).
\end{eqnarray}

\subsection{Free scalar field}

Energy density of the free scalar field has the form
\begin{equation}
\rho =  \frac{1}{2}\frac{\pi^2_{{\phi}}}{\bar{p}^{3} } 
\end{equation}
and the solution of the effective Friedmann equation (effective background equation)
is the following 
\begin{equation}
\bar{p}(t) = \left(A+Bt^2\right)^{1/3} \label{solfreefield}
\end{equation}
where 
\begin{equation}
A = \frac{1}{6}\kappa\pi^2_{\phi}\gamma^2\Delta \ \  , \ \ B = \frac{3}{2}\kappa \pi^2_{\phi}.
\end{equation}
Solution (\ref{solfreefield}) represents non-singular bouncing evolution and is discussed 
in Ref. \cite{Mielczarek:2008zv}.

For further applications, it will be useful to relate the coordinate time with the conformal 
one $d\eta = dt/a(t)$. Assuming that $\eta(t=0)=0$, we obtain 
\begin{equation}
\eta(t)=\frac{t}{A^{1/6}}   {_2F_1} \left[\frac{1}{2},\frac{1}{6},\frac{3}{2};-\frac{B}{A} t^2\right]. \label{etaFSF}
\end{equation}  

\subsection{Scalar field with a multi-fluid potential}

One can shown \cite{Mielczarek:2008qw} that the restriction $p_{\phi}=w\rho_{\phi}$, where $w=$ const
in the framework of effective LQC, leads to the potential in the form 
\begin{equation}
V(\phi) = \frac{1}{2}\rho_c (1-w)  \frac{1}{\cosh^2\left[\sqrt{6\pi G (1+w)}\phi\right] }.  
\end{equation}
Solution of the equations of motion with this potential has simple analytic form
\begin{equation}
\bar{p}(t) = \bar{p}_c \left( 1+6\pi G \rho_c (1+w)^2 t^2 \right)^{\frac{2}{3(1+w)}}.
\end{equation}
It is worth to mention that for $w=1$ and taking 
\begin{equation}
\bar{p}_c^3=A=\frac{1}{6}\kappa\pi^2_{\phi}\gamma^2\Delta
\end{equation}
we recover the solution (\ref{solfreefield}).

In analogy with the free field case we derive
\begin{equation}
\eta(t)=\frac{t}{\sqrt{\bar{p}_c}}   {_2F_1} 
\left[\frac{1}{2},\frac{1}{3(1+w)},\frac{3}{2};-6\pi G \rho_c (1+w)^2t^2\right].
\end{equation}  

\section{Analytical considerations} \label{sec:Analytic}

The theory of cosmological creation of particles bases on 
idea of ``freezing'' of the vacuum fluctuations.  On the 
mathematical level this process can be seen as a 
squeezing and displacement of the vacuum state $|0\rangle$. 
This is equivalent with the creation of particles.
For the non-interacting field theories the wave function is a product of the 
functions for the particular modes. Therefore the degree of  
squeezing and coherence can be different for the particular modes 
and is determined by the cosmological evolution. The typical 
scale for which squeezing and displacement of the vacuum becomes 
important is the Hubble scale. Modes of quantum fluctuations becomes 
classical (are described be the coherent states) when crossing the Hubble radius. 

To describe process of particles creation quantitatively one can consider 
Bogolyubov transformation between initial and final
states. Then computing the so-called  Bogolyubov coefficients the number 
of produced particles can be obtained. However on the super-horizontal
scales one can in principle obtain $\omega^2_k=k^2+m^2_{\text{eff}}< 0 $ 
and the interpretation in terms of particles fails. Then quantum 
state cannot be interpreted in terms of particles. 
Therefore, and from other reasons, it is useful to consider a correlation function
which is well defined quantity for all energy scales. 
The correlation function for the tensor modes takes the form 
\begin{eqnarray}
\langle 0|  \hat{h}^a_b({\bf x},\eta) \hat{h}^b_a ({\bf y},\eta)| 0 \rangle &=&
4 \frac{16\pi G}{a^2} \int \frac{d^3{\bf k}}{(2\pi)^3} |u(k,\eta)|^2 e^{-i{\bf k \cdot r}}  \nonumber  \\ 
 &=& \int \frac{dk}{k} \mathcal{P}_T(k,\eta) \frac{\sin kr}{kr},  
\end{eqnarray}
where we have defined the power spectrum
\begin{equation}
\mathcal{P}_T(k,\eta) = \frac{64\pi G}{a^2} \frac{k^3}{2\pi^2}  |u(k,\eta)|^2. \label{PowerSpectDef}
\end{equation}
The power spectrum can be related later to the amplitude of the CMB fluctuations.
Therefore it is crucial to determinate this function.  

Another way to describe physical properties of the quantum state is the mentioned 
method of Bogolyubov coefficients. The relation between annihilation and creation operators 
for the initial and for the final state is given by the Bogolyubov transformation      
\begin{eqnarray}
\hat{b}_{{\bf k}} &=& B_{+}(k) \hat{a}_{{\bf k}} + B_{-}(k)^{*}  \hat{a}_{-{\bf k}}^{\dagger} \ , \label{Bog1}  \\
\hat{b}_{{\bf k}}^{\dagger} &=& B_{+}(k)^{*}\hat{a}_{{\bf k}}^{\dagger} + B_{-}(k)\hat{a}_{-{\bf k}}\ , \label{Bog2} 
\end{eqnarray}
where $|B_{+}|^2-|B_{-}|^2=1$. Since we are working in the Heisenberg description the vacuum state does not 
change during the evolution. It results that
$\hat{b}_{{\bf k}}| 0_{\text{in}}\rangle=B_{-}(k)^{*}  \hat{a}_{-{\bf k}}^{\dagger}| 0_{\text{in}}\rangle $  is 
different from zero when $B_{-}(k)^{*}$ is a nonzero function. This means that in the final state  
the graviton field considered is no more in the vacuum state without particles. The number of produced particles in the 
final state is given by  
\begin{equation}
n_{{\bf k}} = \frac{1}{2} \langle 0_{\text{in}} |\left[ \hat{b}_{{\bf k}}^{\dagger}\hat{b}_{{\bf k}}+
 \hat{b}_{-{\bf k}}^{\dagger}\hat{b}_{-{\bf k}} \right]| 0_{\text{in}} \rangle =|B_{-}(k)|^2. \label{particles}
\end{equation}
The energy density of gravitons is given by   
\begin{equation}
d\rho_{\text{gw}} = 2 \cdot \hslash \omega \cdot  \frac{4 \pi \omega^2   d\omega}{(2\pi c)^3} \cdot |B_{-}(k)|^2.
\end{equation}
where we used definition (\ref{particles}). To describe the spectrum of gravitons it 
is common to use the parameter
\begin{equation}
\Omega_{\text{gw}}(\nu) =\frac{\nu}{\rho_*}\frac{d \rho_{\text{gw}}}{d \nu}
\label{omegaGW}
\end{equation}
where $\rho_{\text{gw}} $ is the energy density of gravitational waves 
and $\rho_*$ is the present critical energy density.

\subsection{Free scalar field}

\begin{figure}[ht!]
\centering
\includegraphics[width=8cm,angle=0]{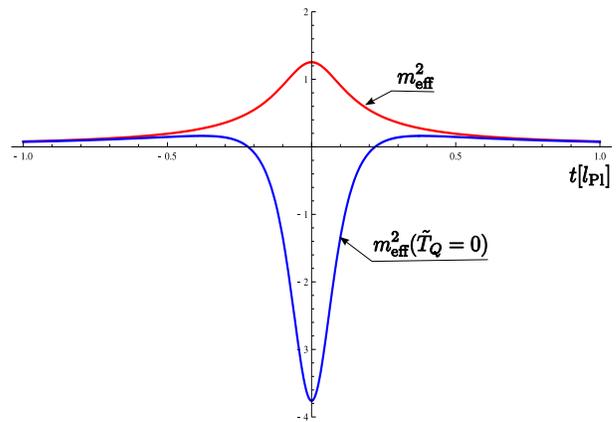}
\caption{Evolution of the effective masses $m^2_{\text{eff}}$ and $m^2_{\text{eff}}(\tilde{T}_Q=0)$.
In this figure we have assumed $\pi_{\phi}=0.1\ l_{\text{Pl}}$}
\label{MeffFree}
\end{figure}

Based on solution (\ref{solfreefield}) we derive
\begin{equation}
m^2_{\text{eff}}=\frac{\kappa^2\pi^4_{\phi}}{4} 
\frac{\left(t^2+\frac{1}{9}\gamma^2\Delta \right)}{\left(A+Bt^2\right)^{5/3}} \geq 0. 
\end{equation}
We show this function in Fig. \ref{MeffFree}.
We compare it with the classical expression
\begin{equation}
m^2_{\text{eff}}(\tilde{T}_Q=0)=\frac{\kappa^2\pi^4_{\phi}}{4} 
\frac{\left(t^2-\frac{1}{3}\gamma^2\Delta \right)}{\left(A+Bt^2\right)^{5/3}}. 
\end{equation}
The difference is significant since now  effective mass is a non-negative function, $m^2_{\text{eff}} \geq 0$.
One can also compare this with the case when the source term corrections were neglected. Then, as 
can be found in Ref. \cite{Mielczarek:2008pf}, the  effective mass is negative in some regime 
and behaves like the classical one. Here difference is crucial and has important 
consequences. Namely since $m^2_{\text{eff}} \geq 0$ we have $\omega_k^2\geq 0$ and interpretation in 
terms of particles is possible on all scales. This becomes a nice property also when the Hamiltonian 
of the perturbations is minimized to find a proper vacuum state.  It can be shown that 
when $\omega_k^2\geq 0$, the Hamiltonian has a minimum for all $k$ and a well defined vacuum can be found.
Otherwise for some $k<k_{\text{x}}$, the lowest-energy instantaneous vacuum state does not exist.
 
\begin{figure}[ht!]
\centering
\includegraphics[width=8cm,angle=0]{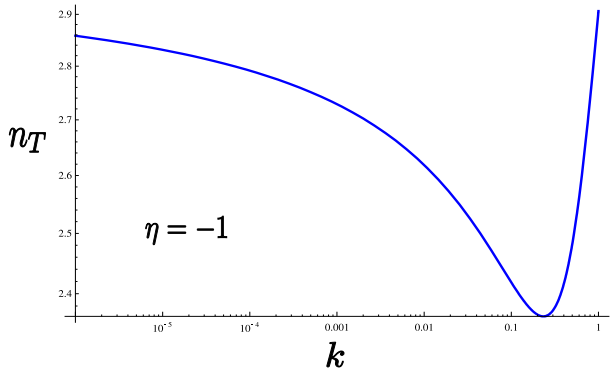}
\caption{Running spectral index on the super-horizontal scales.}
\label{SpectIndex}
\end{figure}

Now we are going to consider the pre-bounce limit. Taking $|t|\rightarrow \infty$, we find 
\begin{equation}
m^2_{\text{eff}} \rightarrow \frac{1}{4} \frac{1}{\eta^2}.
\end{equation}
The normalised solution of the equation (\ref{modeeq}) has the form
\begin{equation}
u(k,\eta) =  \sqrt{\frac{\pi}{2}} e^{i\pi/4} \frac{1}{\sqrt{2k}} \sqrt{-\eta k} H^{(1)}_{0}(-\eta k). \label{ModesSol1}
\end{equation}
We have chosen here advanced modes and performed normalisation with use of the Wronskian condition.
In the super-horizontal limit $-\eta k \ll 1$ we can apply the approximation
\begin{equation}
H^{(1)}_{0}(x) \simeq 1+i\frac{2}{\pi}\left[\ln \left( \frac{x}{2} \right) +\gamma_E  \right], \label{H0expansion}
\end{equation}
where $\gamma_E=0.57721\dots$ is Euler-Mascheroni constant. 
Expression for the power spectrum in the super-horizontal limit is therefore
\begin{equation}
\mathcal{P}_T(k)=\mathcal{A} k^3\left\{ 1+\frac{4}{\pi^2}\left[\ln\left(-\frac{k\eta}{2}\right)+\gamma_E\right]^2 \right\}, 
\label{PTFreefield}
\end{equation}
where 
\begin{equation}
\mathcal{A}=4\sqrt{\frac{2}{\pi}} \left(\frac{3}{2}\right)^{1/6}\left(\frac{l_{\text{Pl}}}{\pi_{\phi}}\right).
\end{equation}
To investigate $k$ dependence in formula (\ref{PTFreefield}) we define the spectral index 
\begin{equation}
n_T = \frac{d \ln \mathcal{P}_T(k)}{d \ln k}
\end{equation}
and obtain
\begin{equation}
n_T = 3+\frac{8}{\pi^2}\frac{\ln\left(-\frac{k\eta}{2}\right)+\gamma_E }
{1+\frac{4}{\pi^2}\left[\ln\left(-\frac{k\eta}{2}\right)+\gamma_E\right]^2 }.
\end{equation}
We show this function for some fixed time in Fig. \ref{SpectIndex}. 
We find that the resulting spectral index is blue and approaching $n_T = 3$ for $k\rightarrow 0$.
This blue-tilted spectrum was predicted earlier in \cite{Mielczarek:2008pf}.  
Recent investigations suggest that also for inflationary cosmology with holonomy corrections 
obtained spectrum is blue-tilted and $n_T = 3$ at super-horizontal scales \cite{Grain:2009kw}.  

\subsection{Multi-fluid potential}

Now we are going to perform similar analysis for the model with a multi-fluid potential.
We obtain formula
\begin{eqnarray}
m^2_{\text{eff}} &=& \bar{p}_c \frac{1}{8} \kappa^2 \rho_c^2(1+w)^2(3w-1)(1+6\pi G \rho_c(1+w)^2t^2 )^{\alpha} \times   
\nonumber    \\ 
                &\times& \left\{t^2-
\frac{4}{3} \frac{\Delta\gamma^2}{(1+w)(3w-1)}\left[1-\frac{2}{3}\frac{(1+3w)}{(1+w)}  \right]  \right\},  
\end{eqnarray}
where
\begin{equation}
\alpha = -\frac{2}{3} \frac{2+3w}{1+w}. 
\end{equation}
In the limit $|t|\rightarrow \infty$ we obtain
\begin{equation}
m^2_{\text{eff}} \rightarrow \frac{6w-2}{(1+3w)^2} \frac{1}{\eta^2},
\end{equation}
where we changed time to conformal. Advanced and normalised solution of the equation (\ref{modeeq}) in
the considered limit is
\begin{equation}
u(k,\eta) = \sqrt{-k\eta} \sqrt{\frac{\pi}{4k}} e^{i\frac{\pi}{2}\left(|\nu|+\frac{1}{2}\right)} H^{(1)}_{|\nu|}(-\eta k) 
\end{equation}
where
\begin{equation}
\nu^2=\frac{9}{4}\frac{(1-w)^2}{(1+3w)^2}. \label{nu2mfp}
\end{equation}
Power spectrum of the perturbations is then given as 
\begin{equation}
\mathcal{P}_{T}(k) \propto (-k\eta)^{3-2|\nu|},
\end{equation}
where super-horizontal approximation
\begin{equation}
H^{(1)}_{n}(x) \simeq -\frac{i}{\pi} \Gamma(n) \left(\frac{x}{2}\right)^{-n} \ \ \text{for} \ \ x \ll 1
\end{equation}
has been used. It must be stressed that the above formula does not hold for $n=0$ (w=1).
In that case another expansion (\ref{H0expansion}) must be applied. 

It is worth to mention that scale invariant spectrum $|\nu|=\frac{3}{2}$ is recovered both
for $w=-1$ and $w=0$, as it can be directly seen from (\ref{nu2mfp}). 
This duality was investigated in Ref. \cite{Wands:1998yp} in context of 
the free scalar field perturbations.

\subsection{Sub-horizontal solutions} 

Since now we were only concerned with the pre-bounce phase. Now we are going to 
evolve modes through the bounce. We firstly consider the case of  
modes which stay under the Hubble radius before the bounce. For that 
modes the initial vacuum state is given by 
\begin{equation}
u_{\text{in}} = \frac{e^{-ik\eta}}{\sqrt{2k}}. \label{SubHorSol1}
\end{equation}
This can be obtained as a limit of the mode function (\ref{ModesSol1}) for $-k\eta \gg 1$.
To be specific, let us consider the model with a free scalar field and $\pi_{\phi}=0.1 l_{\text{Pl}}$.

\begin{figure}[ht!]
\centering
\includegraphics[width=8cm,angle=0]{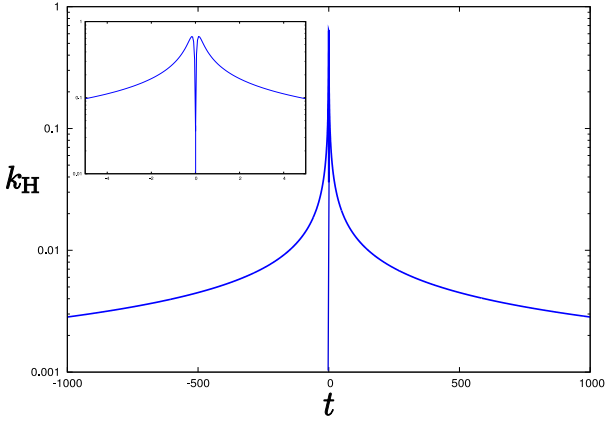}
\caption{Evolution of the Hubble wave number $k_{\text{H}}= a|H|$. Here $\pi_{\phi}=0.1 l_{\text{Pl}}$.}
\label{Hubble}
\end{figure}

At the Hubble radius we have
\begin{equation}
k_{\text{H}}= a|H|
\end{equation}
which is shown in Fig. \ref{Hubble}.
We see that for initial time, let us say $t=-1000 l_{\text{Pl}}$, all modes with $k>0.003$ are 
well described by the function (\ref{SubHorSol1}). This solutions however do not 
hold during the phase of bounce. Close to the bounce one can approximate 
\begin{equation}
m^2_{\text{eff}} \approx m^2_{\text{eff}}(t=0) = 
\frac{1}{(54)^{1/3}} \kappa \left( \pi_{\phi} \rho_c \right)^{2/3} \equiv k^2_0.
\end{equation}
For the considered conditions we obtain $k_0 \simeq 1.12$. In this approximation the solutions 
during the bounce phase are 
 \begin{equation}
u_{\text{bounce}} = \frac{A_k}{\sqrt{2\Omega}}e^{-i\Omega\eta}+\frac{B_k}{\sqrt{2\Omega}}e^{i\Omega\eta},
\end{equation}
where $\Omega=\sqrt{k^2+k^2_0}$. Finally, in the post-bounce phase we have a 
superposition of advanced and retarded modes 
\begin{equation}
u_{\text{out}} = \frac{\alpha_k}{\sqrt{2k}}e^{-ik\eta}+\frac{\beta_k}{\sqrt{2k}}e^{ik\eta}.
\end{equation}
Here the relation $|\alpha_k|^2-|\beta_k|^2=1$ holds, as a consequence of the normalisation condition.
Now we have to match solutions from the three considered regions to determinate 
coefficients $\alpha_k$ and $\beta_k$. In order to do that we must specify a time 
when the matching is performed. We choose it in the mirror points $-t_{-}=t_{+}$ 
where $H^2$ reaches its maximal value. Then $-t_{-}=t_{+}=t_{0}$ where 
\begin{equation}
t_0 = \frac{1}{\sqrt{24\pi G \rho_c}}.
\end{equation} 
With use of equation (\ref{etaFSF}) we obtain 
\begin{equation}
\eta_0 = \eta(t_0)=\frac{{_2F_1} \left[\frac{1}{2},\frac{1}{6},\frac{3}{2};-1\right]}{\sqrt{3\kappa}\rho_c^{1/3}
\left(\pi^2_{\phi}/2\right)^{1/6}}.
\end{equation}
For the considered setup we obtain $\eta_0\simeq 0.285$.

\begin{figure}[ht!]
\centering
\includegraphics[width=8cm,angle=0]{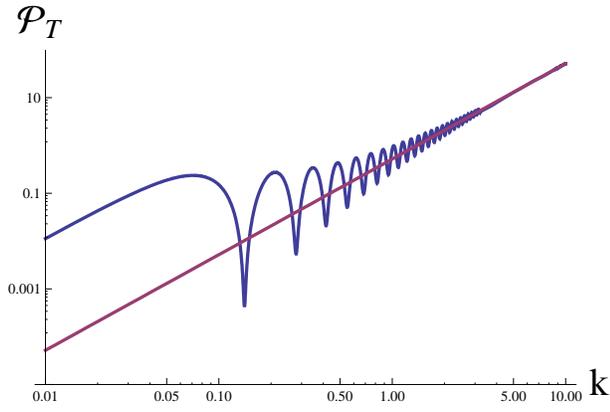}
\caption{Oscillating tensor power spectrum of the sub-horizontal modes at $t=50\ l_{\text{Pl}}$. 
Straight line represents spectrum of reference $\mathcal{P}_T \propto k^2$.}
\label{SpectrumSubHor}
\end{figure}

In order to derive formulas for the coefficients $\alpha_k$ and $\beta_k$, we define the matrices 
\begin{eqnarray}
{\bf M_0} &=& \left(\begin{array}{cc}  \frac{e^{-ik\eta_-}}{\sqrt{2k}}  & \frac{e^{ik\eta_-}}{\sqrt{2k}} \\  
-i\sqrt{\frac{k}{2}} e^{-ik\eta_-}  &  i\sqrt{\frac{k}{2}} e^{ik\eta_-} \end{array}  \right),    \\
{\bf M_1} &=&  \left(\begin{array}{cc} \frac{e^{-i\Omega\eta_-}}{\sqrt{2\Omega}}  & 
\frac{e^{i\Omega\eta_-}}{\sqrt{2\Omega}} \\  
 -i\sqrt{\frac{\Omega}{2}} e^{-i\Omega\eta_-}  &  i\sqrt{\frac{\Omega}{2}} e^{i\Omega\eta_-} \end{array}  \right),    \\
{\bf M_2} &=&  \left(\begin{array}{cc} \frac{e^{-i\Omega\eta_+}}{\sqrt{2\Omega}}  & 
\frac{e^{i\Omega\eta_+}}{\sqrt{2\Omega}} \\  
-i\sqrt{\frac{\Omega}{2}}e^{-i\Omega\eta_+}  & i\sqrt{\frac{\Omega}{2}} e^{i\Omega\eta_+} \end{array}  \right),   \\
{\bf M_3} &=&  \left(\begin{array}{cc}  \frac{e^{-ik\eta_+}}{\sqrt{2k}}  & \frac{e^{ik\eta_+}}{\sqrt{2k}} \\  
-i\sqrt{\frac{k}{2}} e^{-ik\eta_+}  &  i\sqrt{\frac{k}{2}} e^{ik\eta_+} \end{array}  \right).    
\end{eqnarray}
Then matching conditions can be economically written as 
\begin{equation}
\left(\begin{array}{c} \alpha_k  \\ \beta_k \end{array}  \right) = 
{\bf M_3}^{-1}{\bf M_2}{\bf M_1}^{-1}{\bf M_0}\left(\begin{array}{c} 1 \\ 0 \end{array}  \right).
\end{equation}
Multiplying these matrices we obtain
\begin{eqnarray}
\alpha_k &=& \frac{-i \cos(2\eta_0 k)+\sin(2 \eta_0 k)}{2 k \Omega}\times \nonumber \\
       &\times& \left[2 i k \Omega \cos(2 \eta_0 \Omega)+\left(k^2+\Omega^2\right) \sin(2\eta_0 \Omega)\right] \\
\beta_k &=& -\frac{i \left(k^2-\Omega^2\right) \sin(2\eta_0 \Omega)}{2 k \Omega}
\end{eqnarray}
The resulting square of the amplitude for the out state modes is 
\begin{widetext}
\begin{equation}
|u_{\text{out}}|^2=\frac{\left(k^2+\Omega ^2\right)^2-k_0^4 \cos[4 \eta_0 \Omega ]-k^2_0\sin[2 \eta_0 \Omega ] \left((k+\Omega )^2 \sin[2k(\eta-\eta_0)+2\Omega\eta_0]-(k-\Omega )^2 \sin[2k(\eta-\eta_0)-2\Omega\eta_0]\right) } 
{8 k^3 (k^2+k_0^2)}. \label{Subhoramp}
\end{equation}
\end{widetext}
Based on this result one can calculate power spectrum of perturbations. We show this spectrum 
in Fig. \ref{SpectrumSubHor}.
Obtained spectrum exhibits sub-horizontal oscillations. This effect can be intuitively 
understood when analogy with Schr\"{o}dinger equation is employed. Namely the mode equations 
are equivalent to a one dimensional Schr\"{o}dinger equation with potential $V=-m^2_{\text{eff}}$. 
Here spatial variable is replaced by the conformal time $\eta$.
In the employed approximation potential is square well of width $2\eta_0$ and depth $m^2_{\text{eff}}(t=0)$.
Therefore the evolution of the given mode can be seen as transition of a particle over the potential well. 
Amplifications of the amplitude of transmission correspond to resonances between the width of the potential
and the phase shift.

\begin{figure}[ht!]
\centering
\includegraphics[width=8cm,angle=0]{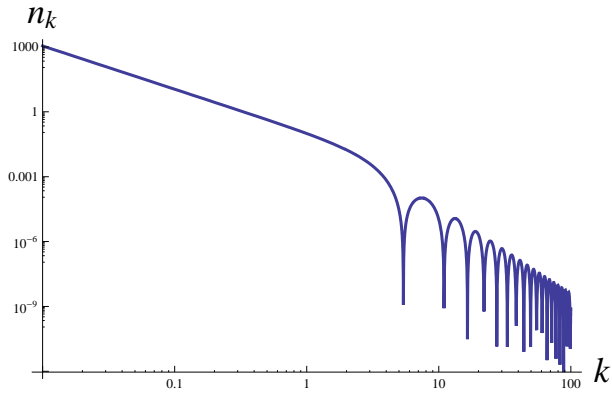}
\caption{Occupation number of the gravitons in the post-bounce state.}
\label{Occupation number}
\end{figure}

\begin{figure}[ht!]
\centering
\includegraphics[width=8cm,angle=0]{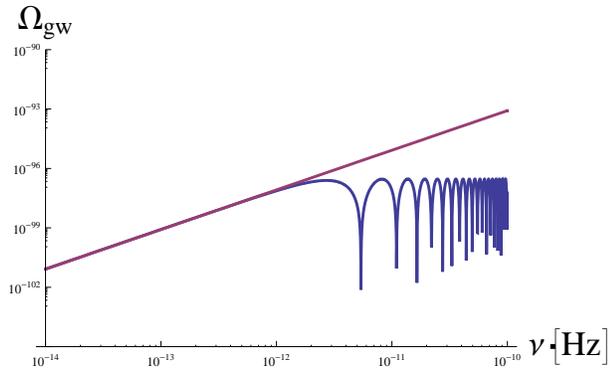}
\caption{Parameter $\Omega_{\text{gw}}$ (bottom curve). 
Straight line represents low energy approximation $\Omega_{\text{gw}} \propto \nu^{-2}$.}
\label{Omega}
\end{figure}

It can be shown that the obtained coefficients $(\alpha_k,\beta_k)$ are in fact the Bogolyubov coefficients
$\alpha_k=B_+$ and $\beta_k=B_-$. Therefore the number of produced gravitons is given by 
\begin{equation}
n_k = |\beta_k|^2= \frac{k_0^4\sin^2\left(2\eta_0 \sqrt{k^2+k^2_0}\right)}{k^2(k_0^2+k^2)}.
\end{equation}
We show this dependence in Fig. \ref{Occupation number}.
Now it is straightforward to calculate the parameter $\Omega_{\text{gw}}$. 
We show this function in Fig. \ref{Omega}.
We compare it with the obtained low energy approximation $\Omega_{\text{gw}} \propto \nu^{-2}$. 
Obtained values of $\Omega_{\text{gw}}$ are many orders of magnitude below the present 
threshold for detection. The results obtained were performed for some simplified model and for 
the fixed value of $\pi_{\phi}$. In particular in the low energy limit 
$\Omega_{\text{gw}} \sim \pi_{\phi}^{2/3}$, therefore effect of varying $\pi_{\phi}$ is considerable. 
However, we do not expect significant changes due to the approximations performed. This statement will be 
confirmed by the numerical simulations in Section \ref{sec:Numerical}.

\subsection{Super-horizontal solutions}

In the previous subsection we have shown solutions of the mode equation 
(\ref{modeeq}) in the sub-horizontal limit. 
Now we are going to study the super-horizontal $k\rightarrow 0$ limit. 

We introduce new variable in the form   
\begin{equation}
f =\sqrt{a} u,
\end{equation}
and change the conformal time to the coordinate one $dt=ad\eta$. 
Then equation (\ref{modeeq}) can be rewritten in the form 
\begin{equation}
\frac{d^2f}{dt^2}+\Omega^2(k,t)f=0. \label{EqOmega2}
\end{equation}
Here the parameter $\Omega^2(k,t)$ is defined as follows
\begin{equation}
\Omega^2(k,t)=\left( \frac{k}{a}\right)^2+\epsilon\frac{\tilde{T}_Q}{a^2}
-\frac{3}{2}\left(\frac{\ddot{a}}{a}+\frac{1}{2}H^2\right). 
\end{equation}
We have introduced here parameter the $\epsilon$ to trace effects 
of the holonomy corrections in the later equations. In the classical limit
we should take $\epsilon=0$ while in presence of the holonomy corrections  $\epsilon=1$.
Taking $k=0$ and introducing new complex variable $z \in \mathbb{C}$,
\begin{equation}
z=\frac{1}{2}+i\frac{1}{2}\sqrt{6\pi G \rho_c}(1+w) t
\end{equation}
we can rewrite equation (\ref{EqOmega2}) in the following form 
\begin{equation}
\frac{d^2f}{dz^2}+Q(z)f = 0 \label{EqQ}
\end{equation}
where
\begin{equation}
Q(z)=\frac{\alpha_2z^2+\alpha_1z+\alpha_0}{z^2(z-1)^2}.
\end{equation}
The coefficients are
\begin{eqnarray}
\alpha_0 &=& \frac{9(1+2w)-4\epsilon(1+3w)}{36(1+w)^2},  \\     
\alpha_1 &=& -\frac{w}{(1+w)^2}, \\
\alpha_2 &=& \frac{w}{(1+w)^2} 
\end{eqnarray}
and it will be useful later to remember that $\alpha_1+\alpha_2=0$.
Now introducing the new variable 
\begin{equation}
f(z)=z^{L}(z-1)^{K}g(z)
\end{equation}
with 
\begin{eqnarray}
L &=& \frac{c}{2} \\
K &=& \frac{a+b+1-c}{2}
\end{eqnarray}
one can rewrite equation (\ref{EqQ}) as a hypergeometric equation
\begin{equation}
z(1-z)\frac{d^2g}{dz^2}+[c-(a+b+1)z]\frac{dg}{dz}-abg=0.
\end{equation}
Solution of this equation is given by the hypergeometric functions
\begin{equation}
g(z)= C {_2F_1(a,b,c;z)}.
\end{equation}
Furthermore, we have a system of equations for the coefficients 
\begin{eqnarray}
\alpha_0+L(L-1) &=& 0, \\
\alpha_1+ab-2KL-2L(L-1)&=& 0, \\
\alpha_2-ab+2KL+L(L-1)+K(K-1)&=& 0. 
\end{eqnarray}

One can find that, since $\alpha_1+\alpha_2=0$, we have either $K=L$
or $K=1-L$, where $L=\frac{1}{2}\left(1\pm \sqrt{1-4\alpha_0}\right)$.
For $K=L$ we find
\begin{eqnarray}
a &=& \frac{1}{2} \left( 2c-1 \pm \sqrt{1+4\alpha_1}\right), \\
b &=& 2c-a-1, \\
c &=& 2L = 1\pm \sqrt{1-4\alpha_0}, 
\end{eqnarray}
and while $K=1-L$ we have
\begin{eqnarray}
a &=& \frac{1}{2} \left(1 \pm \sqrt{1+4\alpha_1}\right), \\
b &=& 1-a, \\
c &=& 2L = 1\pm \sqrt{1-4\alpha_0}. 
\end{eqnarray}

As an exemplary solution we consider $w=1$ case both classically ($\epsilon=0$) and with holonomy corrections 
to the mode equation ($\epsilon=1$).
Then  since $\alpha_1=-\alpha_2=-1/4$ we have 
\begin{equation}
a=b=c-\frac{1}{2}
\end{equation}
where 
\begin{equation}
c_{\pm}(\epsilon=1)=1\pm\frac{\sqrt{47}}{6} \ \text{and} \ c_{\pm}(\epsilon=0)= 1\pm\frac{\sqrt{7}}{2}
\end{equation}

\begin{figure}[ht!]
\centering
\includegraphics[width=8cm,angle=0]{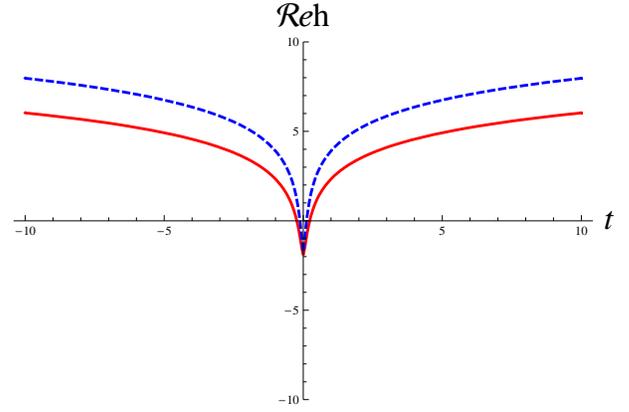}
\caption{Real components of $h$ variable. Dashed (blue) line represents solution 
with $c_{+}(\epsilon=1)$ while straight (red) line represents solution with $c_{+}(\epsilon=0)$.}
\label{SHRe}
\end{figure}

In Fig. \ref{SHRe} we show solutions for the real components of $h$ variable.

\begin{figure}[ht!]
\centering
\includegraphics[width=8cm,angle=0]{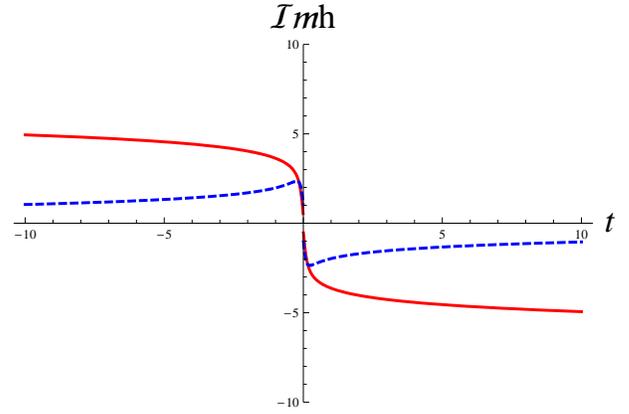}
\caption{Imaginary components of $h$ variable. Dashed (blue) line represents solution 
with $c_{+}(\epsilon=1)$ while straight (red) line represents solution with $c_{+}(\epsilon=0)$.}
\label{SHIm}
\end{figure}

In Fig. \ref{SHIm} we show solutions for the imaginary components of the $h$ variable.

\begin{figure}[ht!]
\centering
\includegraphics[width=8cm,angle=0]{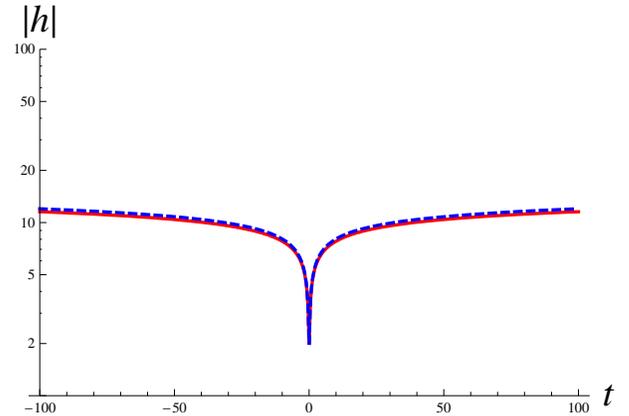}
\caption{Absolute values of $h$ variable. Dashed (blue) line represents solution 
with $c_{+}(\epsilon=1)$ while straight (red) line represents solution with $c_{+}(\epsilon=0)$.}
\label{SHAbs}
\end{figure}

In Fig. \ref{SHAbs} we show solutions for the absolute value of the $h$ variable.

As it can be seen, solutions with and without the quantum holonomy corrections are qualitatively similar.
Another observations is that for the times $t \gg 1$ evolution takes a logarithmic form. This result 
can be understood considering equation (\ref{modeeq}) in the classical limit and taking $k\rightarrow 0$.
Then one can find approximate solution in the form
\begin{equation}
h \simeq A_k+B_k \int^{\eta} \frac{d\eta'}{a^2(\eta')} 
\end{equation} 
where $A_k$ and $B_k$ are some constants. For the models considered in the present paper
we have $a \propto (\pm \eta)^{2/(1+3w)}$  and $\pm\eta \propto (\pm t)^{\frac{1+3w}{3(1+w)}}$.
Here we have $+$ sign for the expanding phase and $-$ sign for the contracting one. Therefore 
for the considered $w=1$ case we find 
\begin{equation}
h \simeq \tilde{A}_k+\tilde{B}_k \ln(\pm t) \ \ \text{for} \ \ |t| \gg 1
\end{equation} 
in agreement with the solutions found in this subsection.

\section{Numerical investigations} \label{sec:Numerical}

As it was shown in the previous section, analytic solutions of the mode equation 
are available only in some limits. Namely for both $t^2\rightarrow \infty$ and  $k\rightarrow 0$.
Also for $k\rightarrow \infty$ an approximate solution was found. 
It is however not sufficient to describe whole spectrum of the gravitational waves 
produced on the bounce phase since the interesting intermediate regimes are 
unexplored. Therefore numerical analysis is required.   

In the numerical computations we are going to solve the autonomous system 
of equations 
\begin{eqnarray}
\frac{du}{d\eta} &=& \pi_u, \\
\frac{d\pi_u}{d\eta} &=&- \left[ k^2+ m^2_{\text{eff}}(t)\right] u, \\ 
\frac{dt}{d\eta} &=& a(t),
\end{eqnarray}
where $a(t)$ and $ m^2_{\text{eff}}(t)$ are defined for particular 
background dynamics. In the considered models with free scalar field
and multi-fluid potential these functions are given by analytical 
expressions. Since canonical variables $u,\pi_u \in \mathbb{C}$
we decompose
\begin{eqnarray}
u     &=& u_1+iu_2,      \\
\pi_u &=& \pi_{u1}+i\pi_{u2}.
\end{eqnarray}
Now it is crucial to define proper initial conditions 
for $(u_1,u_2,\pi_{u1},\pi_{u2})$ for some time $\eta_0$.
It is always unambiguous how to choose a proper vacuum defined 
on the cosmological background. However on the sub-horizontal
scales, when Minkowski space approximation holds, we can set
\begin{eqnarray}
u_1(\eta_0) &=& \frac{1}{\sqrt{2k}}\cos(k\eta_0), \\ 
u_2(\eta_0) &=& -\frac{1}{\sqrt{2k}}\sin(k\eta_0)
\end{eqnarray}
and
\begin{eqnarray}
\pi_{u1}(\eta_0) &=& -\sqrt{\frac{k}{2}}\sin(k\eta_0), \\ 
\pi_{u2}(\eta_0) &=& -\sqrt{\frac{k}{2}}\cos(k\eta_0)
\end{eqnarray}
at some time $\eta_0$. Here we set initial values like in the model 
of sub-horizontal modes studied in the previous section. Therefore 
analysis is correct for the modes with with $k>0.003$. 
 
\begin{figure}[ht!]
\centering
\includegraphics[width=8cm,angle=0]{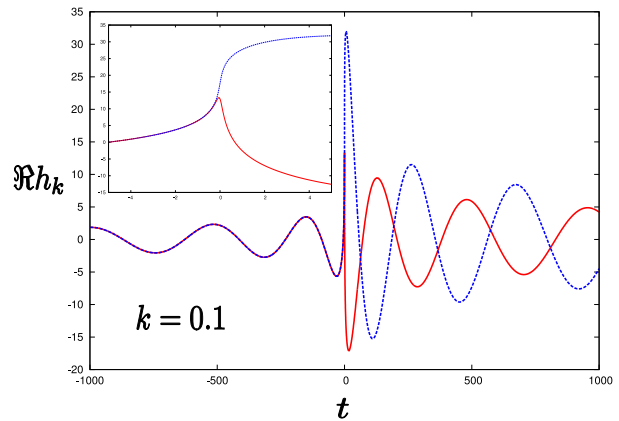}
\caption{Evolution of the modes with $k=0.1$. Dotted (blue) curve represent solution of mode
equations with holonomy corrections. Straight (red) curve represent solution of mode equations 
without holonomy effects.}
\label{k01evol}
\end{figure}

\begin{figure}[ht!]
\centering
\includegraphics[width=8cm,angle=0]{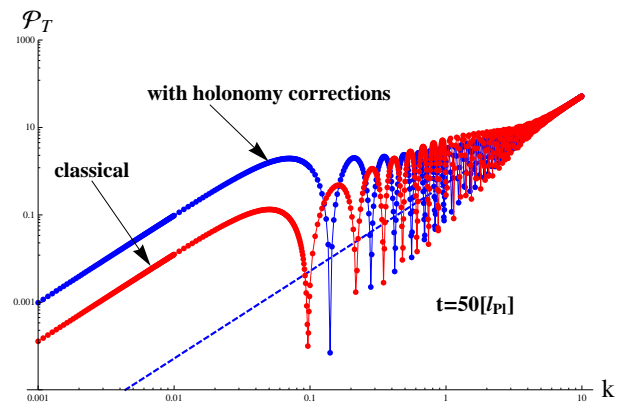}
\caption{Comparison between sub-horizontal spectra obtained with and without 
holonomy corrections to mode equation. Dashed (blue) line represents rescaled initial 
vacuum power spectrum.  
}
\label{SpectA}
\end{figure}

In Fig. \ref{k01evol} we plot evolution of the $k=0.1$ mode during the bounce phase.
We compare here the evolution of modes with and without holonomy corrections to the modes equation.
We see that close to the turning point the effects of the holonomy corrections become significant.
However the further oscillating evolution does not change qualitatively. The difference is 
some suppression of the amplitude of perturbations due to the quantum corrections.  
This feature can be also seen in Fig. \ref{SpectA} where classical and quantum corrected tensor power spectra
are shown.
We find, comparing with the classical case, that quantum holonomy effects amplify low energy modes. 
Therefore tensor power spectra increases by about one order of magnitude. For the high energies classical
spectra starts to dominate slightly. It is also worth to notice that oscillations do not overlap.

To impose initial conditions on the super-horizontal scales one can use 
instantaneous vacuum. This is however possible only for values of 
$k$ fulfilling $\omega^2_k \geq 0$. As we have found earlier
this condition is fulfilled for all $k$ in the model with the free scalar field.
Therefore initial instantaneous vacuum state can be defined on all length scales.
It can be shown that Hamiltonian of perturbations at time $\eta_0$ is minimised 
for  
\begin{eqnarray}
u(\eta_0) &=& \frac{1}{\sqrt{2\omega_k}} \\ 
\pi_{u}(\eta_0) &=& -i\sqrt{\frac{\omega_k}{2}}
\end{eqnarray}
where $\omega_k=\sqrt{k^2+m^2_{\text{eff}}}$. 

\begin{figure}[ht!]
\centering
\includegraphics[width=8cm,angle=0]{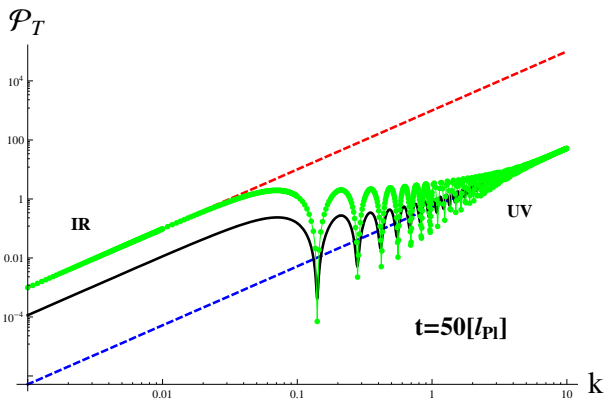}
\caption{Whole post-bounce tensor spectra in 
presence of the holonomy corrections. Green 
points comes from the numerical simulations.
Black line is the analytical spectrum from 
the model given by Eq. \ref{Subhoramp}. 
Dashed (red and blue) lines represents UV 
and IR behaviours, in both cases $\mathcal{P}_T \propto k^2$.
}
\label{IRUV}
\end{figure}

In Fig. \ref{IRUV}  we show tensor power spectrum at post-bounce stage 
($t=50\ l_{\text{Pl}}$) with imposed instantaneous vacuum initial conditions 
at $t=-1000\ l_{\text{Pl}}$.
The characteristic feature of the spectrum are oscillations. Moreover 
both UV and IR behaviours are in the form $\mathcal{P}_T \propto k^2$.
We see that analytical model given by Eq. \ref{Subhoramp} fairly good 
overlap with the numerical results. Especially structure of oscillations
is exactly recovered. Also the asymptotic behaviours are consistent. 
The evident discrepancy is the difference in the total amplitude. In fact 
this difference can be suitably adjusted varying parameters of the model 
$\eta_0$ and $k_0$. Then low energy behaviour can be exactly recovered.
However it introduces additional phase shift and 
structures of oscillation no longer overlap. It is also important to 
note that effect of the imposed instantaneous vacuum initial conditions
is negligible in the range studied. Therefore Minkowski vacuum approximation
is still valid. 

The obtained power spectrum can be now applied as an initial condition for the 
inflationary modes. We expect that super-horizontal part of the spectrum 
does not change during the inflationary phase. However the UV part becomes 
nearly flat (depending on the model of inflation). It is possible that 
the oscillating features will also survive giving the footprints of the 
bouncing phase. However the further analysis has to be performed to 
approve these speculations. In particular inflationary power spectrum
with the obtained bouncing initial conditions must be calculated. 
Then it will be possible to compute the B-type polarization spectra of CMB.
Therefore a way to relate the quantum cosmological effects with the 
low energy physics becomes potentially available.

\section{Summary} \label{sec:Summary}

In this paper we have considered influence of Loop Quantum Gravity effects on 
the gravitational waves propagation in the flat FRW cosmological background.
The considerations presented based on the semi-classical approach where 
quantum effects are introduced by corrections to the classical equations of motion.
This approach was successfully applied to the homogeneous models. In this case
good agreement between results of fully quantum and semi-classical analyses
was found. Here we have applied semi-classical approach to the inhomogeneous model 
where inhomogeneity is treated perturbatively. Therefore perturbations had no 
influence on background. In general both classical and quantum back-reaction 
effects can be important close to the phase of bounce. Here we assumed that 
they can be neglected. We have also not considered effects of the quantum 
fluctuations of background on the inhomogeneities. Quantum effects were 
introduced by the so called holonomy corrections. In the homogeneous models these corrections 
lead to absence of the initial singularity and emergence of the bounce phase. 
Effects of the other known type of LQG corrections, the inverse volume ones, 
were studied earlier in numerous papers. Here we considered self-consistent 
model where holonomy effects influence both background and perturbations (gravitational waves). 
In the earlier studies effects on background and perturbations were studied 
independently. In particular in Ref. \cite{Mielczarek:2008pf} a model of gravitational
waves production during the holonomy-inducted bounce phase was investigated. 
In Ref. \cite{Barrau:2008hi,Grain:2009kw} effects of holonomy corrections 
on the gravitational waves in inflationary phase were studied. However 
quantum effects on the background dynamics were neglected there. Moreover 
quantum-corrected source term was not taken into account in those studies. 
Linear part of this term vanish in the classical limit. However, its contribute 
while holonomy corrections are present. Therefore source term has to be taken into account 
in the full treatment. In the present paper we have included effects of this term.

We have considered models with both free scalar field and self-interacting field
with multi-fluid potential. In both cases scalar field is a monotonic function and
can be treated as a internal time variable.

We have shown that in the model with the free field, effective mass term $m^2_{\text{eff}}$
for gravitational waves is a non-negative function. This is not the case for the models
with multi-fluid potential. 
We have found solutions of the mode function in the pre-bounce phase and 
determined the power spectra of the obtained perturbations.
Then we have considered sub-horizontal solutions during the bounce phase.
We matched the solutions from pre-bounce, bounce and post-bounce phases.
Based on this we have found power spectrum of gravitational waves 
and determined Bogolyubov coefficients. Then number of produced gravitons
$n_{\bf k}$ and the parameter $\Omega_{\text{gw}}$ were calculated. We have found
that $\Omega_{\text{gw}}$ reaches  $10^{-96}$ which is far below any 
observational bounds. These results were obtained for fixed parameter 
$\pi_{\phi}=0.1 \ l_{\text{Pl}}$. 

Based on analytical considerations we have found that power spectrum 
exhibits oscillations on sub-horizontal scales. An intuitive explanation 
of this effect was given. We have also solved the model analytically in 
the super-horizontal limit. These results indicate that quantum corrections 
do not introduce qualitative difference in the power spectrum on these scales.
Therefore the obtained lack of power on the large scales is a feature of the bouncing 
evolution and not of the quantum corrections to the mode equation. 

Subsequently we have investigated the model numerically. We have approved 
presence of the oscillations emerged from the simplified analytical considerations. 
Both numerical and analytical results were compared. We have found good 
qualitative and quantitative agreement. We have also approved earlier observation 
that quantum corrections does not introduce qualitative difference in the power spectrum. 
The only differences observed were in total amplitude and phase of oscillations.  

Imposing initial instantaneous vacuum state we have also studied the low energy
part of the power spectrum. Therefore we have finally found the full shape of the 
tensor power spectrum. This spectrum can be used to study further phenomenological 
consequences. In particular, it can be applied as an initial condition for the
inflationary modes. Then we expect that the sub-horizontal part of the spectrum
becomes flat while super-horizontal form survive. It is also possible that 
sub-horizontal oscillations survive as features of the dominant nearly flat inflationary
spectrum. Therefore two observational effects of the bouncing phase can be 
distinguished: oscillations and lack of power on the super-horizontal scales. 
These effects can potentially be tested with the future CMB missions like 
Planck \cite{:2006uk} or proposed CMBPol \cite{Baumann:2008aq}. Especially promising 
are observations of the CMB polarization. Here bounce can lead to the low multipoles 
suppression in the B-type spectrum. At present projects like Clover \cite{North:2008jq}, 
QUaD \cite{Castro:2009ej} or QUIET \cite{Samtleben:2008sj} are aiming 
to detect this spectrum and first results are expected in the near future. Therefore it is 
the next step to derive quantitative predictions of the CMB features from the 
presented model.       

\begin{acknowledgments}
Author is grateful to Francesca Vidotto for discussion during the conference 
"Quantum Gravity in Cracow$^2$" 19-21 XII 2008, Poland,  where part of these results were
presented.
\end{acknowledgments}

\appendix 
\section{Loop Quantum Gravity with gravitational waves} \label{Appendix1}

Loop Quantum Gravity (LQG) describes the gravitational field as $SU(2)$ 
non-Abelian gauge field using background independent methods. 
The canonical fields are so called Ashtekar variables 
$(A=A^i_a\tau_i dx^a,E=E^a_i\tau^i \partial_a)$ which take value in $\mathfrak{su}(2)$
and $\mathfrak{su}(2)^*$ algebras respectively and they fulfil the Poisson bracket
\begin{equation}
\{A^i_a({\bf x}),E^b_j({\bf y})\} = \gamma \kappa \delta^b_a \delta^i_j \delta^{(3)}({\bf x-y})
\end{equation}
where $\kappa = 8\pi G$ and  $\gamma$ is the Barbero-Immirzi parameter.
These variables are analogues of the vector potential and the electric field
in electrodynamics. The Ashtekar variables are related
with triad representation. In LQG gauge fields describe only 
spatial part $\Sigma$ when time is treated separately.

In cosmological applications we perturb basic variables around a background 
\begin{eqnarray}
E^a_i &=& \bar{E}^a_i +\delta E^a_i,    \\
A^i_a &=& \bar{A}^i_a +\delta A^i_a.  
\end{eqnarray}
For the spatially flat FRW background components have the following form
\begin{eqnarray}
\bar{E}^a_i &=& \bar{p} \delta^a_i,   \\
\bar{A}^i_a &=&\gamma\bar{k}\delta^i_a, 
\end{eqnarray}
where  $\bar{p}=a^2$ and $ \bar{k} =  \dot{\bar{p}}/2\bar{p}$.
Perturbations can be split for the scalar, vector and tensor parts.
For the purpose of this paper we consider here only the gravitational waves
(tensor part). Tensor perturbations of the flat FRW metric are introduced as follows
\begin{eqnarray}
g_{00} &=& -N^2+q_{ab}N^aN^b = -\bar{N} = -a^2, \nonumber   \\ 
g_{0a} &=& q_{ab}N^b = 0,   \nonumber            \\
g_{ab} &=& q_{ab} = a^2 [\delta_{ab}+h_{ab}], \nonumber
\end{eqnarray}
with the conditions $h^a_a=\partial_a h^a_b=0$ and $|h_{ab}|\ll 1$. In the TT
gauge $h^1_1=-h^2_2=h_{\oplus}$ and $h^1_2=h^2_1=h_{\otimes}$.

Now we are going to perturb  the Hamiltonian of the theory. 
The full Hamiltonian is composed of the gravitational and $H_{\text{G}}$ and
matter $H_{\text{m}}$ parts. 
Hamiltonian $H_{\text{G}}$ takes the form of a liner combination of the constraints
\begin{equation}
H_{\text{G}}= \int_{\Sigma} d^3 {\bf x} \, (N^iG_i+N^aC_a+NS). \nonumber
\end{equation}
Spatial diffeomorphisms constraint:
\begin{equation}
C_a = E^b_i F^i_{ab} - (1-\gamma^2)K^i_a G_i.  \nonumber
\end{equation}
Gauss constraint:
\begin{equation}
G_i = D_a E^a_i =\partial_aE^a_i+\epsilon_{ijk} A^j_a E^a_k.   \nonumber
\end{equation}
Scalar constraint:
\begin{equation}
 S=\frac{E^a_iE^b_j}{\sqrt{|\det E|}} \left[ {\varepsilon^{ij}}_k F_{ab}^k -
 2(1+\gamma^2)K^i_{[a} K^j_{b]} \right]   \nonumber
\end{equation}
where $F=dA +\frac{1}{2}[A,A]$. However thanks to the quantum gravity effect 
this Hamiltonian undergoes modifications. These modifications can be introduced
on the phenomenological level by the replacement 
\begin{equation}
\bar{k} \rightarrow \frac{\sin n \bar{\mu}\gamma \bar{k}}{n \bar{\mu}\gamma}
\end{equation}
in the classical expressions. Here 
\begin{equation}
\bar{\mu}=\sqrt{\frac{\Delta}{\bar{p}}} \ \ \text{where}\ \ \Delta = 2\sqrt{3}\pi \gamma l^2_{\text{Pl}}.  \nonumber
\end{equation}
This kind of corrections we call holonomy ones. Factor  $n$ can be fixed 
from requirement of the anomaly cancellation \cite{Bojowald:2007cd,Bojowald:2007hv}.
Effective second order Hamiltonian with holonomy corrections takes the form
\begin{eqnarray}
H_G^{\text{phen}} &=& \frac{1}{16\pi G} \int_{\Sigma} d^3x \bar{N} \left[ -6 \sqrt{\bar{p}} 
  \left(\frac{ \sin \bar{\mu}\gamma \bar{k}}{\bar{\mu} \gamma}\right)^2   \right.  \nonumber \\
 &-& \frac{1}{2\bar{p}^{3/2}} \left(\frac{ \sin \bar{\mu}\gamma \bar{k}}{\bar{\mu} \gamma}\right)^2  
(\delta E^c_j \delta E^d_k \delta^k_c \delta^j_d)  \nonumber \\
  &+& \sqrt{\bar{p}}(\delta K^j_c\delta K^k_d \delta^c_k \delta^d_j )  \nonumber \\
  &-&  \frac{2}{ \sqrt{\bar{p}} } 
\left(\frac{\sin 2\bar{\mu}\gamma \bar{k}}{2\bar{\mu} \gamma}\right)
 (\delta E^c_j \delta K^j_c) \nonumber \\
&+& \left.  \frac{1}{\bar{p}^{3/2}} (\delta_{cd}\delta^{jk} \delta^{ef} \partial_e E^c_j \partial_f E^d_k)  \right]  
\end{eqnarray}
where for tensor modes 
\begin{eqnarray}
\delta E^a_i &=& -\frac{1}{2}\bar{p}h^a_i \label{deltaEApp} \\
\delta K^i_a &=& \frac{1}{2}\left[ \dot{h}^i_a+ 
\left( \frac{\sin 2\bar{\mu}\gamma \bar{k} }{ 2\bar{\mu}\gamma} \right)h^i_a\right]
\end{eqnarray}
Based on the Hamilton equations 
\begin{eqnarray}
\delta \dot{E}^a_i &=& \left\{ \delta E^a_i, H_G^{\text{phen}}+H_m \right\},       \\
\delta \dot{K}^i_a &=& \left\{ \delta K^i_a, H_G^{\text{phen}}+H_m \right\},       
\end{eqnarray}
 we obtain equation
\begin{equation}
\ddot{h}^i_a + 2\bar{k}\dot{h}^i_a-\nabla^2h^i_a+
 T_Qh^i_a =  16 \pi G \Pi^i_{Qa}
\end{equation}
where 
\begin{eqnarray}
T_Q &=&  -2\left(\frac{\bar{p}}{\bar{\mu}} \frac{\partial \bar{\mu}}{\partial \bar{p}}  \right)
\bar{\mu}^2\gamma^2\left( \frac{\sin \bar{\mu}\gamma \bar{k} }{ \bar{\mu}\gamma} \right)^4, \label{TQApp}       \\
\Pi^i_{Qa} &=& \left[ \frac{1}{3V_0} \frac{\partial \bar{H}_{\text{m}}}{\partial \bar{p}} 
\left( \frac{\delta E^c_j\delta^j_a \delta^i_c}{\bar{p}} \right)\cos 2\bar{\mu}\gamma \bar{k}
+\frac{\delta H_m}{\delta(\delta E^a_i)}  \right] \label{PiQApp} 
\end{eqnarray}
are quantum holonomy corrections.

We consider homogeneous scalar field with the Hamiltonian
\begin{equation}
{H}_{m}=\int_{\Sigma} d^3{\bf x} \, \bar{N} \left( \frac{1}{2}\frac{\pi^2_{{\phi}}}{\sqrt{|\det E|}}  
  + \sqrt{|\det E|} V({\phi}) \right), 
\end{equation}
where  up to the second order 
\begin{eqnarray}
\sqrt{\det E } &=&  \bar{p}^{\frac{3}{2}} \left[1 +\frac{1}{2\bar{p}}\delta^i_a \delta E^a_i 
- \frac{1}{4\bar{p}^2} \delta^i_a \delta E^a_j \delta^j_b \delta E^b_i  \right. \nonumber    \\
  &+& \left. \frac{1}{8\bar{p}^2} \delta^i_a \delta E^a_i  \delta^j_b \delta E^b_j \right],     \\
\frac{1}{\sqrt{\det E }} &=& \frac{1}{\bar{p}^{\frac{3}{2}}} \left[1 - \frac{1}{2\bar{p}}\delta^i_a \delta E^a_i 
+ \frac{1}{4\bar{p}^2} \delta^i_a \delta E^a_j \delta^j_b \delta E^b_i \right. \nonumber    \\
 &+& \left. \frac{1}{8\bar{p}^2} \delta^i_a \delta E^a_i  \delta^j_b \delta E^b_j \right]. 
\end{eqnarray}
However, since $\delta^{ab}h_{ab} = 0 \Rightarrow  \delta^i_a \delta E^a_i = 0$ the above expansion simplifies. Then
\begin{equation}
{H}_{m}=\bar{H}_{m} + \frac{1}{4}\int_{\Sigma} d^3{\bf x} \, \frac{\bar{N}}{\sqrt{\bar{p}}} 
 \left( \frac{1}{2}\frac{\pi^2_{{\phi}}}{\bar{p}^{3} }-V({\phi}) \right)\delta^i_a \delta E^a_j \delta^j_b \delta E^b_i
 +\mathcal{O}(E^3). \nonumber  
\end{equation}
Now we can derive variation 
\begin{equation}
\frac{\delta H_m}{\delta(\delta E^a_i)} = \frac{1}{2}\frac{\bar{N}}{\sqrt{\bar{p}}} 
 \left( \frac{1}{2}\frac{\pi^2_{{\phi}}}{\bar{p}^{3} }-V({\phi}) \right)\delta^i_b \delta^k_a \delta E^b_k \label{der1App}
\end{equation}
and derivative 
\begin{equation}
\frac{\partial \bar{H}_{\text{m}}}{\partial \bar{p}} = \frac{3}{2}
 V_0\frac{\bar{N}}{\sqrt{\bar{p}}}\left( -\frac{1}{2}\frac{\pi^2_{{\phi}}}{\bar{p}^{3} }+V({\phi}) \right). \label{der2App}
\end{equation}
One can now easily find that in the classical limit, when we set $\cos(2\bar{\mu}\gamma \bar{k})=1$ in expression  
(\ref{PiQApp}), the source term vanish. This is due to the opposite signs of the bracketed expression in 
equations (\ref{der1App}) and (\ref{der2App}). When quantum holonomy corrections are present we have
\begin{equation}
\cos(2\bar{\mu}\gamma \bar{k}) =1-2\frac{\rho}{\rho_c}, 
\end{equation}
which can be found from background equations of motion.
Therefore the form of the quantum corrections simplifies to 
\begin{eqnarray}
T_Q        &=& \frac{8\pi G}{3} \frac{\bar{p}\rho^2}{\rho_c},  \\
\Pi^i_{Qa} &=&  \Pi_Qh^i_a = \frac{1}{2}\bar{p} \frac{\rho}{\rho_c} \left(2V-\rho\right)h^i_a,        
\end{eqnarray}
where we have chosen $\bar{N}=\sqrt{\bar{p}}$ and adopted the expression (\ref{deltaEApp}).

\end{document}